\documentclass[floats,floatfix,showpacs,amssymb,prd,twocolumn,superscriptaddress,nofootinbib,nolongbibliography,reprint]{revtex4-2}

\usepackage{amssymb,amsmath,verbatim,mathtools,needspace,enumitem,etoolbox,graphicx,physics,microtype,afterpage,bigints}
\usepackage{gensymb}

\usepackage[dvipsnames, usenames]{xcolor}
\definecolor{linkcolor}{rgb}{0.0,0.3,0.5}
\definecolor{dodgerblue}{HTML}{1E90FF}
\usepackage[unicode, colorlinks=true, linkcolor=linkcolor, citecolor=linkcolor, filecolor=linkcolor,urlcolor=linkcolor, pdfusetitle]{hyperref}
\usepackage[all]{hypcap}
\usepackage[T1]{fontenc}
\usepackage[utf8]{inputenc}
\usepackage{tabularx}

\newcommand{\ssim}{\mathchar"5218\relax\,}

\renewcommand{\vec}[1]{\mathbf{#1}}

\makeatletter
\newcommand*{\balancecolsandclearpage}{\close@column@grid \cleardoublepage \twocolumngrid}
\makeatother

\def\apj{\rm{ApJ}}
\def\apjl{\rm{ApJ}}

\def\mnras{\rm{MNRAS}}
\def\prd{\rm{PRD}}
\def\prl{\rm{PRL}}
\def\prx{\rm{PRX}}

\def\cqg{\rm{CQG}}

\newcommand{\bham}{\affiliation{School of Physics and Astronomy \& Institute for Gravitational Wave Astronomy, \\ University of Birmingham, Birmingham, B15 2TT, United Kingdom}}

\newcommand\orcid[1]{\href{https://orcid.org/#1}{$\!$\includegraphics[scale=0.006]{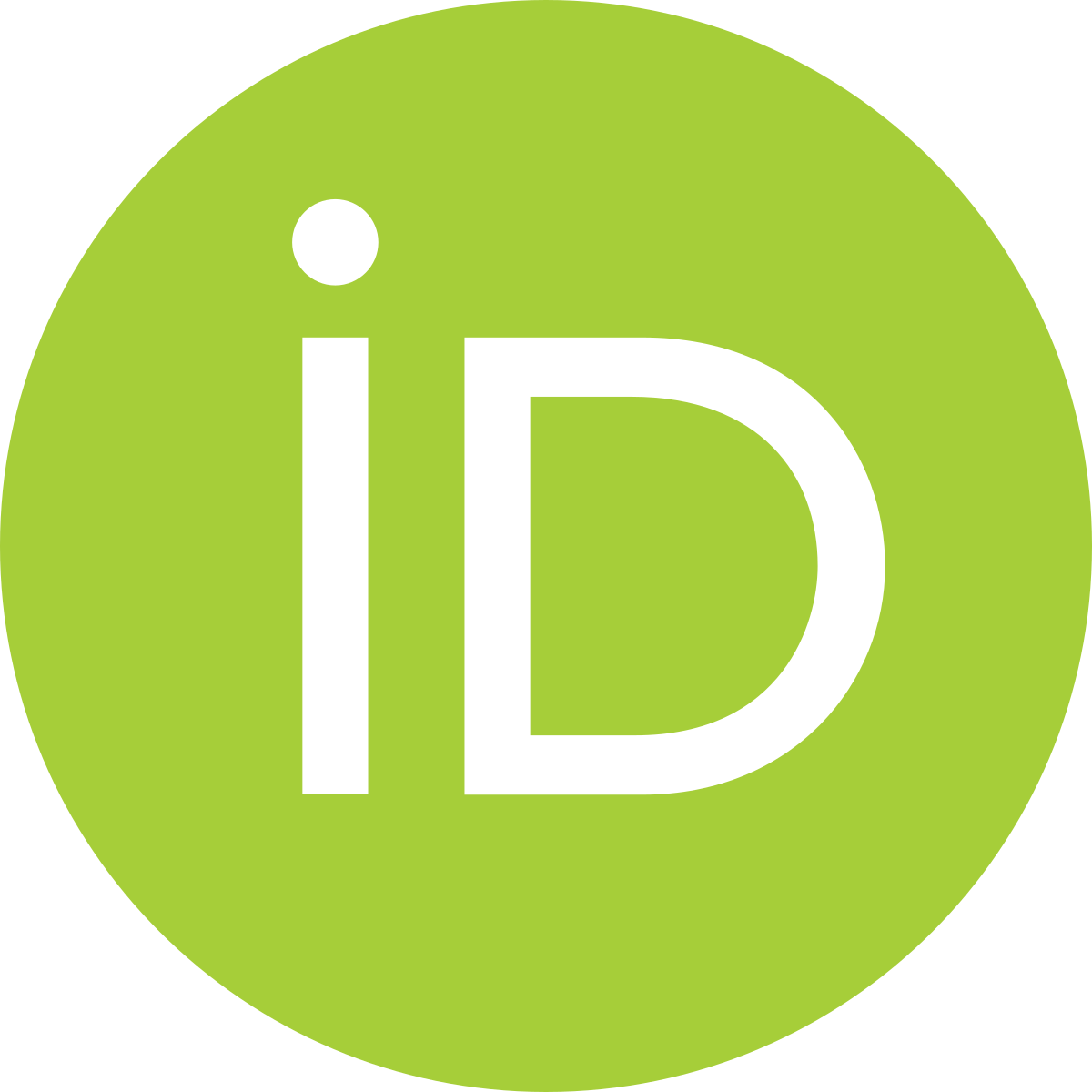} $\!\!$}}

\newcommand{\chieff}{\chi_\mathrm{eff}}
\newcommand{\chip}{\chi_\mathrm{p}}

\begin{document}

\title{A generalized precession parameter $\chi_\mathrm{p}$ to interpret gravitational-wave data}

\author{Davide Gerosa \orcid{0000-0002-0933-3579}}
\email{d.gerosa@bham.ac.uk}

\author{Matthew Mould \orcid{0000-0001-5460-2910}}

\author{Daria Gangardt \orcid{0000-0001-7747-689X}}

\author{Patricia Schmidt \orcid{0000-0003-1542-1791}}

\author{Geraint Pratten \orcid{0000-0003-4984-0775}}

\author{Lucy M. Thomas \orcid{0000-0003-3271-6436}}

  \bham


\date{\today}

\begin{abstract}

Originally designed for waveform approximants, the effective precession parameter $\chip$ is the most commonly used quantity to characterize spin-precession effects in gravitational-wave observations of black-hole binary coalescences. We point out that the current definition of $\chip$ retains some, but not all, variations taking place on the precession timescale. We rectify this inconsistency and propose more general definitions that either fully consider or fully average those oscillations. Our generalized parameter $\chip\in[0,2]$ presents an exclusive region $\chip>1$ that can only be populated by binaries with two precessing spins. We apply our prescriptions to current LIGO/Virgo events and find that posterior distributions of $\chip$ tend to show longer tails at larger values. This appears to be a  generic feature, implying that (i) current $\chip$ measurement errors might be underestimated, but also that (ii) evidence for spin precession in current data might be stronger than previously inferred. Among the gravitational-wave events released to date, that which shows the most striking behavior is GW190521.

\end{abstract}

\maketitle

\section{Introduction}

Spin precession is a key phenomenological feature of black-hole (BH) binary coalescences. As the two BHs inspiral toward merger due to the emission of gravitational-waves (GWs), relativistic spin-spin and spin-orbit couplings cause the orbital plane and the spins to precess about the direction of the total angular momentum \cite{1994PhRvD..49.6274A}.

BH binary spin precession is often characterized using a single effective parameter, denoted as $\chip$. First introduced by \citeauthor{2015PhRvD..91b4043S} \cite{2015PhRvD..91b4043S} for waveform building purposes, $\chip$ is now widely used in state-of-the-art analyses of LIGO/Virgo data to infer the occurrence of spin precession \cite{2016PhRvX...6d1015A,2019PhRvX...9c1040A,2020arXiv201014527A}. A confident measurement of $\chip$ away from zero with significant information gain from the prior is considered a strong indication that orbital-plane precession has been measured. %
Alternatively, Ref.~\cite{2020PhRvD.102d1302F}  proposed to quantify spin precession in terms of the  excess signal-to-noise-ratio (SNR) $\rho_{\rm p}$ of a precessing signal compared to a non-precessing one.
  
Most recently, the parameter   $\chip$  was the main tool used by \citeauthor{2020arXiv201014533T}~\cite{2020arXiv201014533T} to claim that, although evidence for spin precession in individual events is mild to moderate, current data show a much stronger \emph{collective} evidence for precessing spins that emerges at the population level.
This has important consequences for the astrophysical interpretation of BH mergers, most notably in terms of their  formation channel(s). Precessing spins are a key prediction of  BH binaries formed in dense clusters, but might also be present in the case of sources formed in isolation because of, e.g., supernova kicks~\cite{2000ApJ...541..319K,2010CQGra..27k4007M,2013PhRvD..87j4028G,2016ApJ...832L...2R,2018PhRvD..98h4036G,2020arXiv201000078S,2020arXiv201109570C}.

In this paper, we %
 reinvestigate the derivation of $\chip$ and rectify an inconsistency in its current definition \mbox{---namely} that only some, but not all, of the precession-timescale oscillations are averaged. Section~\ref{maths} provides a concrete recipe to either retain all such variations or
properly average them.
The latter approach results in an augmented definition of $\chip$ that varies only on the longer radiation-reaction timescale and includes two-spin effects. Some details of the full averaging procedure are postponed to Appendix~\ref{averimp}. Section~\ref{parex} presents a brief exploration of the parameter space using post-Newtonian (PN) integrations and quantifies the extent to which the current definition of $\chip$ fails to properly capture two-spin effects.  Section~\ref{currentevents} explores the consequences of our findings on current LIGO/Virgo events. The proposed generalization of $\chip$ causes long tails in the posterior distributions,
indicating that evidence for spin precession inferred from current data might be underestimated, while the accuracy may be overestimated.
This is a generic feature, with potential consequences for GW population studies. Finally, in Sec.~\ref{concl} we make our conclusions and discuss future prospects.

\section{How to quantify precession?} \label{maths}

Let us consider a quasicircular BH binary with total mass $M=m_1+m_2$, mass ratio $q=m_2/m_1\leq 1$, spin vectors ${\vec{S}}_{1,2}$, and dimensionless spin magnitudes $\chi_{1,2}$. We employ geometric units $G=c=1$.
The orbit-averaged evolution of the orbital angular momentum $\mathbf{L}$ can be written as
\begin{equation}
\frac{d\vec{L}}{dt} = \frac{d\hat{\vec{L}}}{dt} L + \frac{dL}{dt} \hat{\vec{L}} =
(\vec{\Omega}_L   \times \hat{\vec{L}} ) L +  \frac{dL}{dt} \hat{\vec{L}}\,,
 \end{equation}
 where the first term describes precession and the second term encodes radiation reaction.  The magnitude of the angular momentum $\vec{L}$ is related to the orbital separation $r$ by the Newtonian expression $L/M^2 = (r/M)^{1/2} q/(1+q)^2$.
The precession frequency  $\vec{\Omega}_L$ includes contributions from both spins,
\begin{equation}
 \vec{\Omega}_L   = \Omega_1\chi_1  \hat{\vec{S}}_1 +  \Omega_2\chi_2  \hat{\vec{S}}_2\,,
 \end{equation}
which at next-to-leading order in $M^2/L$ are given by~\cite{2008PhRvD..78d4021R}
\begin{align}
\Omega_1 &= \frac{M^2}{2r^3 (1+q)^2} \left[ 4 + 3q - \frac {3q\chieff} {(1+q)}\frac{M^2}{L} \right]\,, \\
\Omega_2 &= \frac{q M^2}{2r^3 (1+q)^2} \left[ 4q + 3 - \frac {3 q\chieff} {(1+q)}\frac{M^2}{L} \right]\,,
\end{align}
where $\chieff$ is the effective spin~\cite{2001PhRvD..64l4013D,2008PhRvD..78d4021R}
\begin{align}
\chieff = \frac{\chi_1 \hat{\vec S}_1 + q\chi_2 \hat{\vec S}_2 }{1+q}\cdot \hat{\vec L} \,.
\label{chieff}
\end{align}

The amount of orbital-plane precession is thus set by the magnitude
\begin{align}
\left| \frac{d\hat{\vec{L}}}{dt}  \right|^2 &= \left(\Omega_1 \chi_1 |\hat{\vec{S}}_1 \times  \hat{\vec{L}}|\right)^2 +   \left(\Omega_2 \chi_2 |\hat{\vec{S}}_2 \times  \hat{\vec{L}}|\right)^2\notag \\
&+ 2 \Omega_1\Omega_2
\chi_1\chi_2\left(\hat{\vec{S}}_1 \times  \hat{\vec{L}}\right) \cdot  \left(\hat{\vec{S}}_2 \times  \hat{\vec{L}}\right) \,.
\end{align}
We follow common practice and describe the geometry of the systems in terms of the tilt angles $\theta_{1,2}$ and the difference $\Delta\Phi$ between the {phases of the} in-plane components of the two spins.\footnote{{The angle} $\Delta\Phi$ is sometimes indicated as $\phi_{12}$ in LIGO/Virgo analyses and data products.} In symbols, these are
\begin{align}
\label{cos theta1}
\cos{\theta_1} &= \hat{\vec{S}}_1 \cdot\hat{\vec{L}}
\, ,\\
\label{cos theta2}
\qquad \cos{\theta_2} &= \hat{\vec{S}}_2 \cdot\hat{\vec{L}}
\, , \\
\label{cos DeltaPhi}
\cos{\Delta\Phi} &=\frac{\hat{\vec{S}}_1 \cross \hat{\vec{L}}}{|\hat{\vec{S}}_1 \cross \hat{\vec{L}}|} \cdot \frac{\hat{\vec{S}}_2 \cross \hat{\vec{L}}}{|\hat{\vec{S}}_2 \cross \hat{\vec{L}}|}
\, ,
\end{align}
which yields
\begin{align}
\left| \frac{d\hat{\vec{L}}}{dt}  \right|^2 &= \left(\Omega_1 \chi_1 \sin\theta_1 \right)^2 +   \left(\Omega_2 \chi_2 \sin\theta_2 \right)^2
\notag\\
&+ 2 \Omega_1\Omega_2 \chi_1\chi_2\sin\theta_1 \sin\theta_2 \cos\Delta\Phi \,. \label{dhatL2}
\end{align}

The argument made in Ref.~\cite{2015PhRvD..91b4043S} where $\chip$ is first introduced can be recast as follows. The factor $\cos\Delta\Phi$ can (in principle, at least) take values between $-1$ and $+1$. At those extrema one has
\begin{align}
\left| \frac{d\hat{\vec{L}}}{dt}  \right|_\pm &= \left|\Omega_1 \chi_1 \sin\theta_1 \pm \Omega_2 \chi_2 \sin\theta_2 \right| \,.\end{align}
The parameter $\chip$ is defined as the arithmetic mean of  these two contributions normalized by the frequency $\Omega_1$, i.e.
\begin{align}
\chip &\equiv \frac{1}{2\Omega_1}  \left(\left| \frac{d\hat{\vec{L}}}{dt}  \right|_+ + \left| \frac{d\hat{\vec{L}}}{dt}  \right|_- \right)
\notag \\
&= \max\left(\chi_1 \sin\theta_1,\tilde\Omega \chi_2 \sin\theta_2\right)\,,
\label{contributionaverage}
\end{align}
where we introduced the ratio between the spin frequencies
\begin{align}
\tilde\Omega = \frac{\Omega_2}{\Omega_1} = q\frac{4q + 3}{4 + 3q}  - \frac{3\chi_{\rm eff}  q^2 (1-q)}{(4+3q)^2(1+q)} \frac{M^2}{L} + {\mathcal O \left(\frac{M^4}{L^2}\right)}\,.
\label{tildeomega}
\end{align}
To leading order in $M^2/L$, one has
\begin{align}
\chip \simeq \max\left(\chi_1 \sin\theta_1, q\frac{4q + 3}{4 + 3q} \chi_2 \sin\theta_2\right)\,,
\label{usualchip}
\end{align}
which is the expression from Ref.~\cite{2015PhRvD..91b4043S} used in current GW analyses (e.g.~\cite{2016PhRvX...6d1015A,2019PhRvX...9c1040A,2020arXiv201014527A}).

While the simplicity of this procedure is appealing, it is worth pointing out that the three angles $\theta_1$, $\theta_2$ and $\Delta\Phi$ all vary on the same timescale {$t_{\rm pre}\propto (r/M)^{5/2}$}. One is {not} justified to devise a procedure that removes the $\Delta\Phi$ dependence from Eq.~(\ref{dhatL2}) while at the same time retaining $\theta_1$ and $\theta_2$. The definition of $\chip$ given in Eq.~(\ref{usualchip}) is therefore inconsistent because it contains some, but not all, short-timescale variations. %
Let us stress that this is not the case for the other commonly used spin parameter $\chieff$, which is a constant of motion at 2PN \cite{2008PhRvD..78d4021R}.

There are two possible strategies one can pursue: either  retain all the precession-timescale variations, or integrate them out.

If precession-timescale variations are to be retained, one can immediately generalize the definition of $\chip$
as the magnitude of ${d\hat{\vec{L}}}/{dt}$ normalized by $\Omega_1$, i.e.:
\begin{align}
\chip \equiv  \left| \frac{d\hat{\vec{L}}}{dt}  \right| \frac{1}{\Omega_1} &= \bigg[\left(\chi_1 \sin\theta_1 \right)^2 +   \left(\tilde\Omega \chi_2 \sin\theta_2 \right)^2
\notag\\
&+ 2 \tilde\Omega\chi_1\chi_2 \sin\theta_1 \sin\theta_2 \cos\Delta\Phi \bigg]^{1/2}\,.
\label{fullchip}
\end{align}

If one instead wishes to remove those variations, Eq.~(\ref{fullchip}) should be precession averaged in a consistent fashion. Given a suitable quantity $\psi(t)$ that parametrizes the precession cycle (this is analogous to, say, Kepler's mean anomaly for the orbital problem), the precession-averaged value of $\chip$ can be found by evaluating
\begin{align}
\langle\chip \rangle = \frac{\displaystyle \bigintssss \chip(\psi)\left(\frac{d\psi}{dt}\right)^{\!-1} d\psi}{  \displaystyle\bigintssss\left(\frac{d\psi}{dt}\!\right)^{-1} \!\!\!\! d\psi}\,.
\label{fullaverage}
\end{align}
When plugging Eq.~(\ref{fullchip}) into Eq.~(\ref{fullaverage}), one should remember that the angles $\theta_1(\psi)$, $\theta_2(\psi)$, and $\Delta\Phi(\psi)$ all vary on the precession timescale and thus depend (perhaps nontrivially) on $\psi$. On the other hand, the ratio $
\tilde\Omega$ is constant at leading order and presents only long-timescale variations if the first PN correction is included, see Eq.~(\ref{tildeomega}).
Two explicit parametrizations
at 2PN are constructed in Refs.~\cite{2015PhRvL.114h1103K,2015PhRvD..92f4016G, 2017PhRvD..95j4004C}. In particular, the parameter $\psi(t)$ can be chosen to be either the angle
\begin{align}
\cos \varphi'
    =\frac{{\vec{S}}_1 \cdot \left[\left({\vec{S}}_1 \times {\vec{L}} \right) \times \vec{S}_2 + \left({\vec{S}}_2 \times {\vec{L}} \right) \times \vec{S}_2\right]}{|{\vec{S}}_1\times{\vec{S}}_2|\,
        |\left({\vec{S}}_1+ {\vec{S}_2}\right)\times {\vec{L}}|}
\label{varphiprime}
\end{align}
or the magnitude of the total spin
\begin{equation}
S= |\vec{S}_1 +\vec{S}_2|\,.
\end{equation}
A practical implementation for $\psi(t)=S(t)$ is provided in Appendix~\ref{averimp}.

We can now put the approximation of Eqs.~(\ref{contributionaverage}) on more formal grounds. Let us picture a simplified precession cycle where $\vec{S}_1$ and $\vec{S}_2$ precess about $\vec{L}$ with constant velocity and constant opening angles. In this case, one has $d\theta_1/dt = d\theta_2/dt= d^2\Delta\Phi/ dt^2 =0$ and the angle $\Delta\Phi$ itself can be used to parametrize the precession cycle.
With these assumptions, Eq.~(\ref{fullaverage}) simplifies to
\begin{align}
&\langle \chip\rangle \simeq \frac{1}{2 \pi}   \int_0^{2\pi} \chip\,  d\Delta\Phi =\notag \\
&\frac{|\chi_1 \sin\theta_1 - \tilde\Omega \chi_2 \sin\theta_2|}{\pi}\,{\rm E} \!\left[-\frac{4 \tilde\Omega  \chi_1  \chi_2 \sin\theta_1 \sin\theta_2}{(\chi_1 \sin\theta_1 - \tilde\Omega \chi_2 \sin\theta_2)^2}\right]
\notag\\&+ \frac{\chi_1 \sin\theta_1 + \tilde\Omega \chi_2 \sin\theta_2}{\pi}\, {\rm E}\! \left[\frac{4 \tilde\Omega  \chi_1  \chi_2 \sin\theta_1 \sin\theta_2}{(\chi_1 \sin\theta_1 + \tilde\Omega \chi_2 \sin\theta_2)^2}\right]
\label{ellipE}
\, ,
\end{align}
where ${\rm E}(m) = \int_0^{\pi/2}  (1-m \sin^2\!x)^{1/2} dx $ is the complete elliptic integral of {the} second kind.
\begin{figure*}
\includegraphics[width=0.98\textwidth]{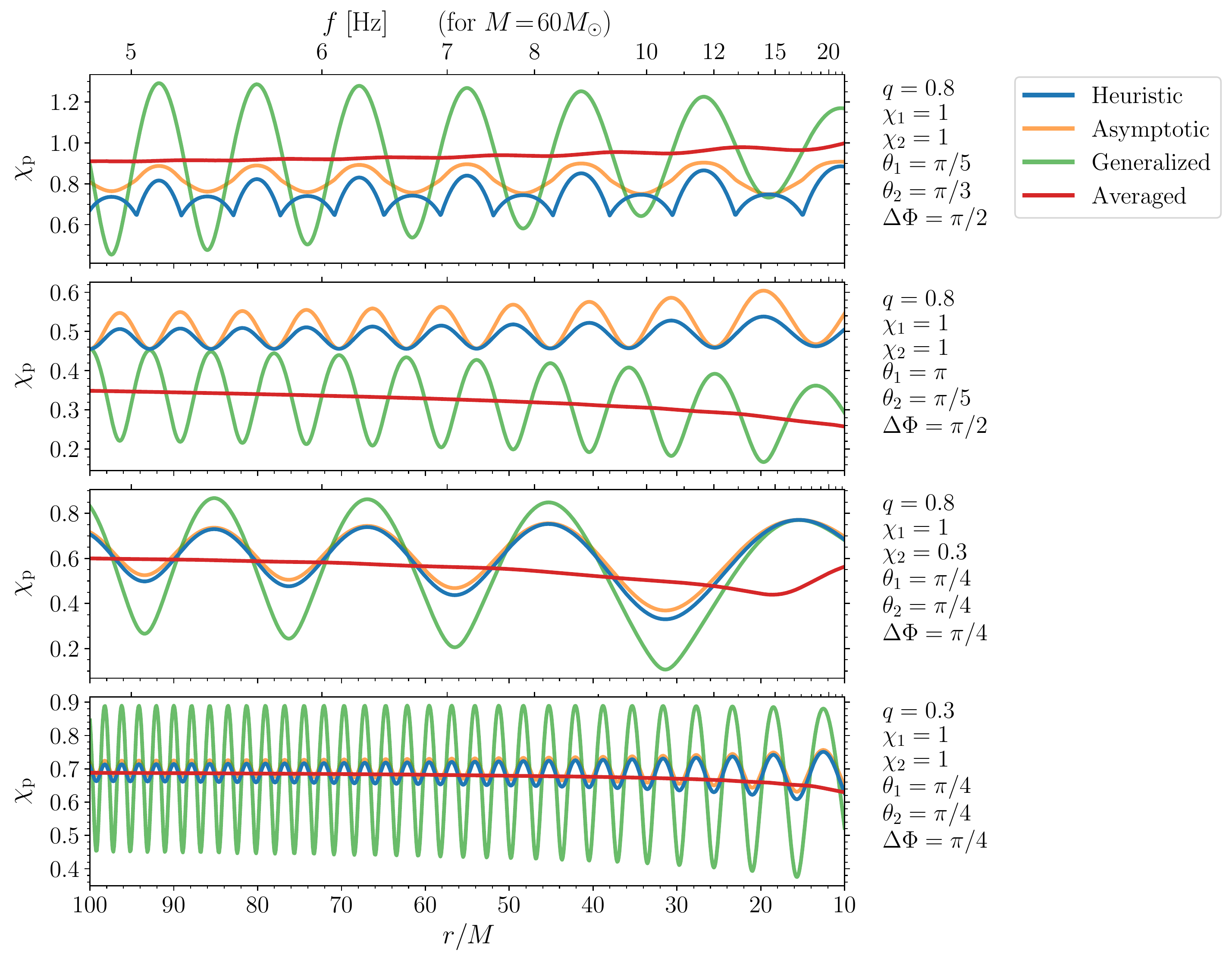}
\caption{Evolution of $\chip$ during the inspiral of four representative BH binaries. The heuristic $\chip$ from  Eq.~(\ref{usualchip}) is shown in blue, the asymptotic $\chip$ from Eq.~(\ref{ellipE}) (here used inappropriately at finite separations) is shown in orange, the generalized $\chip$ from Eq.~(\ref{fullchip}) is shown in green and the averaged $\chip$ from Eq.~(\ref{fullaverage}) is shown red.  Binaries are evolved from  $r =100 M$ to $r=10M$ with initial conditions set by the values indicated to the right of each panel. The top axis indicates the corresponding GW frequency $f =\sqrt{M/\pi^2 r^3}$ for nominal sources with $M=60 M_\odot$.
}
\label{someevols}
\end{figure*}

The conditions $d\theta_1/dt = d\theta_2/dt =d^2\Delta\Phi/ dt^2=0$  %
are satisfied whenever spin-spin couplings can be neglected compared to spin-orbit couplings. This is true at large separations \cite{2015PhRvD..92f4016G} and, indeed, one can show that Eq.~(\ref{ellipE}) is the formal limit of Eqs.~(\ref{fullchip}-\ref{fullaverage}) as $r/M\to\infty$.
At finite separations, however, spin-spin couplings introduce variations of the opening angles (resulting in spin nutations) as well as nonuniform angular velocities of the in-plane spin components.
In particular, geometrical constraints can prevent binaries from ever reaching either $\cos\Delta\Phi = 1$ or $\cos\Delta\Phi = -1$ (the ``librating morphologies'' in the language of Refs.~\cite{2015PhRvL.114h1103K,2015PhRvD..92f4016G}). This is not compatible with the argument made in Eq.~(\ref{contributionaverage}) %
which  relies on $| {d\hat{\vec{L}}}/{dt}  |_\pm$ where $\cos\Delta\Phi=\pm1$.  %
Neglecting nutations can also introduce a significant mismodeling, as the spin angles can vary by as much as  $\Delta\theta_i\sim \pi$ on short timescales~\cite{2016PhRvD..93d4031L,2019CQGra..36j5003G}.

Spin-spin couplings also vanish, trivially, for binaries with a single spin. Let us recall that $S_1\to 0$ corresponds  to $\chi_1\to 0$, while $S_2\to 0$ corresponds to either $\chi_2\to 0 $ or $q\to 0$, which is equivalent to $\tilde\Omega\chi_2\to0$. One can encapsulate both these limits in the quantity 
\begin{align}
\delta\chi = \frac{\min\left(\chi_1 \sin\theta_1,\tilde\Omega \chi_2 \sin\theta_2\right)}{\max\left(\chi_1 \sin\theta_1,\tilde\Omega \chi_2 \sin\theta_2\right)}\,,
\,
\end{align}
such that single-spin binaries corresponds to $\delta\chi = 0$. Equation~(\ref{ellipE}) can now be Taylor-expanded  to obtain
\begin{align}
\langle \chip\rangle \simeq  \max\left(\chi_1 \sin\theta_1,\tilde\Omega \chi_2 \sin\theta_2\right) \left[ 1 + \frac{\delta\chi^2}{4} +\mathcal{O}(\delta\chi^4)\right]\,.
\label{tayloraverage}
\end{align}
The heuristic definition of $\chip$ given in Eq.~(\ref{contributionaverage}) is equivalent to the leading-order term and reduces to it identically if either $S_1=0$ or $S_2=0$. 
The physical scenario where a single spin dominates the precession dynamics was indeed the motivation behind the waveform model developed in Refs.~\cite{2012PhRvD..86j4063S,2014PhRvL.113o1101H,2015PhRvD..91b4043S} where $\chip$ was first introduced. The asymmetric case is explored explicitly in Ref.~\cite{2020PhRvR...2d3096P}.

\begin{figure*}
\includegraphics[width=\textwidth]{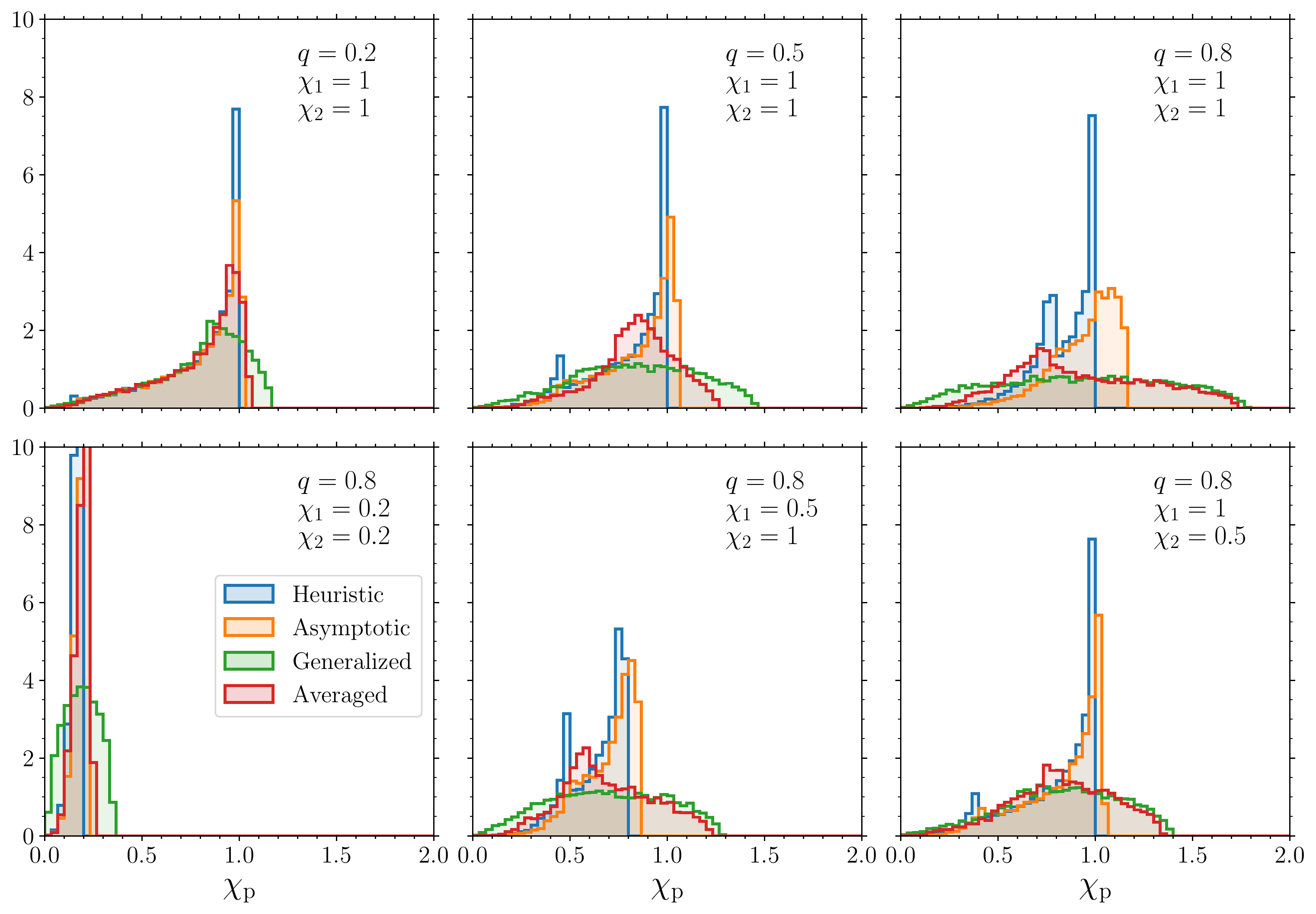}
\caption{Distribution of $\chip$ using the  heuristic [Eq.~(\ref{usualchip}), blue],   asymptotic [Eq.~(\ref{ellipE}), orange], generalized [Eq.~(\ref{fullchip}), green], and averaged [Eq.~(\ref{fullaverage}), red] definitions. Each panel contains a population of $10^4$ sources with fixed values of $q$, $\chi_1$, $\chi_2$ (as indicated in the figure), and isotropic spin directions. Sources are taken at $r\simeq 14 M$, or $f=20$ Hz for a total mass $M=60 M_\odot$.
}
\label{manydistr}
\end{figure*}

 \begin{figure}
\includegraphics[width=\columnwidth]{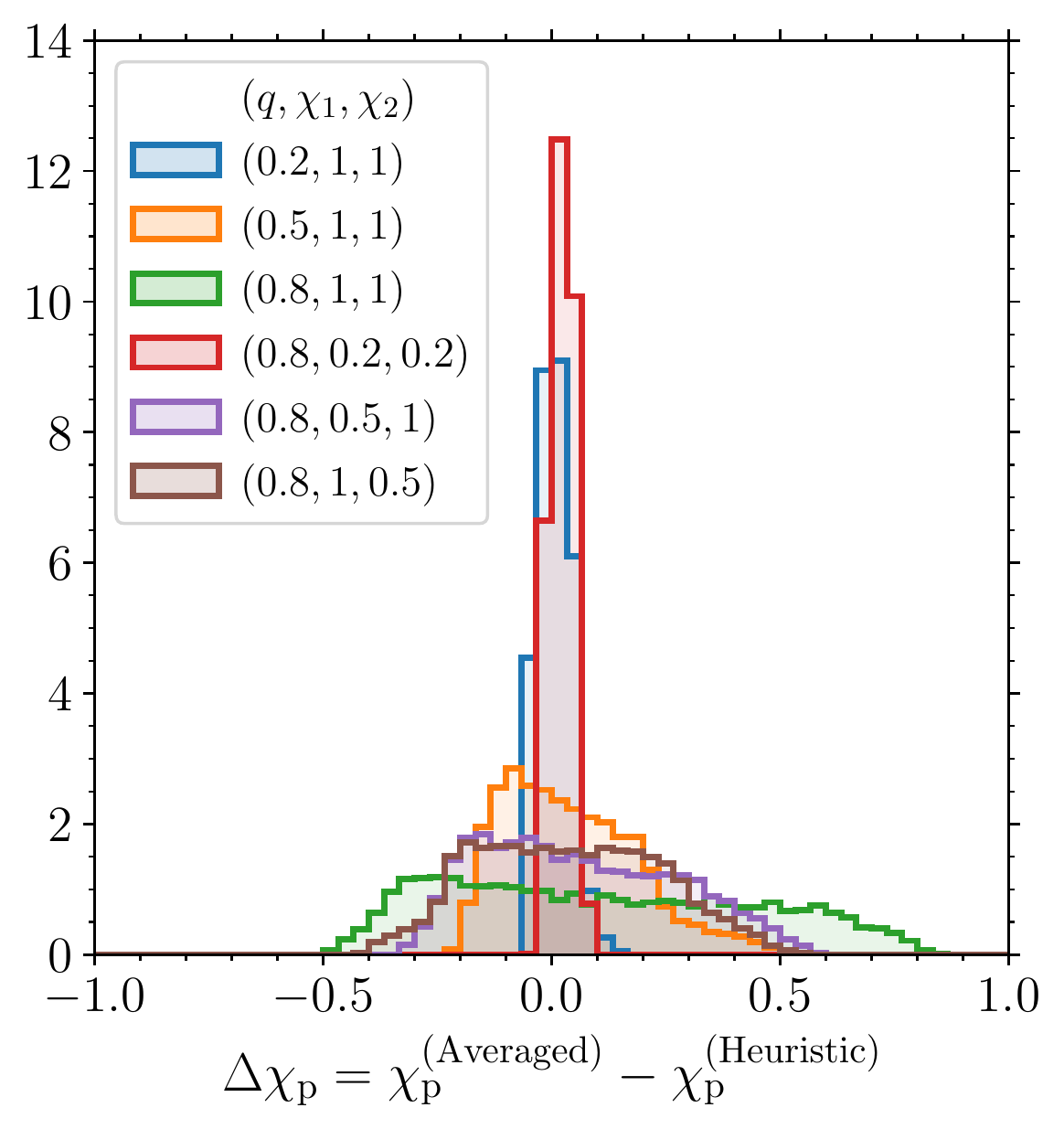}
\caption{
Difference $\Delta\chip$ between the averaged [Eq.~(\ref{fullaverage})] and heuristic [Eq.~(\ref{usualchip})] estimates for the same population of sources shown in Fig.~\ref{manydistr}. Each histogram is produced assuming  isotropic spin directions and fixed values of $q$, $\chi_1$, and $\chi_2$ as indicated in the legend. The parameter $\Delta\chip$ is evaluated at $r\simeq 14M$, equivalent to $f=20$ Hz for $M=60 M_\odot$.
}
\label{diffdistrs}
\end{figure}

\section{Parameter-space exploration}
\label{parex}

We now investigate similarities and differences between the various definitions of $\chip$ using PN integrations.
We use the following terminology:
\begin{align}
{\rm Eq.~(\ref{usualchip})}\;\; &\longrightarrow \;\; {\rm ``Heuristic"}\;\chip, \notag
\\
{\rm Eq.~(\ref{ellipE})}\;\; &\longrightarrow \;\; {\rm ``Asymptotic"}\;\chip, \notag
\\
{\rm Eq.~(\ref{fullchip})}\;\; &\longrightarrow \;\; {\rm ``Generalized"}\;\chip, \notag
\\
{\rm Eq.~(\ref{fullaverage})}\;\; &\longrightarrow \;\; {\rm ``Averaged"}\;\chip. \notag
\end{align}
We assume $\tilde\Omega =  q (4q+3)/(4+3q)$ as in Ref.~\cite{2015PhRvD..91b4043S}. We have verified that the additional PN correction reported in Eq.~(\ref{tildeomega}) is irrelevant.

Figure~\ref{someevols} shows the evolution of $\chi_{\rm p}$ for a set of representative BH binaries. Sources are initialized by specifying values of $q$, $\chi_1$, $\chi_2$, $\theta_1$, $\theta_2$, and $\Delta\Phi$ at $r=100 M$,  and evolved down to $r=10 M$ using the orbit-averaged PN code of Ref.~\cite{2016PhRvD..93l4066G}.

The top two panels of Fig.~\ref{someevols} illustrate cases where the heuristic estimate of $\chip$ (blue) fails to capture the dynamics. The kinks in the topmost panel are due to the 
maximum taken in Eq.~(\ref{usualchip}): for this system, the two terms $\chi_1 \sin\theta_1$ and $\tilde\Omega\chi_2 \sin\theta_2$ alternate their relative importance during each precession cycle, such that selecting only one of them 
introduces sharp features. This specific issue is rectified if one considers the asymptotic expression (orange) which, however, still fails to match either the generalized (green) or the averaged (red) result. This is because the assumptions used to derive Eq.~(\ref{ellipE}) are only valid in the limit $r/M \to \infty$, while here we evaluate the asymptotic $\chip$ inappropriately along the inspiral. The averaged $\chip$ (green) correctly tracks the long-term behavior of
$|{d\hat{\vec{L}}}/{dt}|$. In general, we find that the heuristic estimate of $\chip$ can either overestimate or underestimate the averaged evolution.

Conversely, the bottom two panels of  Fig.~\ref{someevols} present systems where the heuristic expression tracks the overall dynamics more faithfully. Notably, these are  cases where the spin of the primary BH dominates, i.e. closer to the $m_1^2 \chi_1 \gg m_2^2 \chi_2$ limit.  Even in this favorable scenario, however, both the heuristic and the asymptotic definition of $\chip$ retain prominent short-timescale variations which are instead smoothed out by the averaged result. The generalized $\chip$ contains the full precession-timescale dynamics and thus oscillates with even larger amplitude.

A careful inspection of Fig.~\ref{someevols} reveals that the red curve for the averaged evolution of $\chip$ still retains some (much smaller!) variations that correlate with the precession period. This is signaling the breaking down of the timescale separation that underpins the averaging procedure. Much like the quasicircular approximation cannot accurately describe the orbital problem close to merger, averaging over a precession cycle is also less justified at small orbital separations where radiation reaction becomes more  prominent.

Figure~\ref{manydistr} explores the statistical distribution of $\chip$ for all four definitions. Each panel is produced assuming  a population of sources with fixed values of $q$, $\chi_1$, and $\chi_2$, and spin directions distributed isotropically. The values of $\chip$ are evaluated at $r\simeq 14 M$, corresponding to GW frequencies of $f = 20$ Hz for a nominal source with total mass $M=60 M_\odot$ (where we converted frequency to separation using  the Newtonian expression $f =\sqrt{M/\pi^2 r^3}$).

An immediate observation is that the heuristic value of $\chip$ is bounded by $\chip\leq 1$ while the generalized and averaged estimates satisfy $\chip \leq 2$. This is another reflection of the fact that cases where both spins contribute to the precession dynamics cannot be faithfully reduced to a single spin.
The blue histograms for the heuristic  $\chip$ in Fig.~\ref{manydistr}  show two prominent peaks at $\chip =\chi_1$ and $\chip = \chi_2 q (4q+3)/(4+3q)$. These artificial 
features are not present in either the  generalized or the averaged distributions.
Interestingly, the averaged $\chip$ distributions lie between the generalized and the heuristic ones. This is a consequence of the derivation of the heuristic $\chip$ presented in Sec.~\ref{maths} which relies on an inconsistent average.

One could be tempted to evaluate the asymptotic limit reported in Eq.~(\ref{ellipE}) along the inspiral (much as we did in Fig.~\ref{someevols}) because it represents an easy-to-implement, semianalytical expression. The red histograms in Fig.~\ref{manydistr} show that, overall, such an approach would also be inappropriate to describe the precession dynamics. This is because Eq.~(\ref{ellipE})  is only valid in the  $r/M \to \infty$ limit where spin nutations can be neglected and the precessional velocity is approximately constant. It is worth stressing that Eq.~(\ref{ellipE}), and not Eq.~(\ref{usualchip}), provides the correct asymptotic limit of ${\langle \chip \rangle}$ as $r/M \to \infty$. The averaged result agrees with its asymptotic limit  within a few percent only at extremely large separations $r\gtrsim 10^5M$. These values are not accessible by LIGO and Virgo: in the sensitivity windows of the detectors where $r/M$ is of $\mathcal{O}(10)$, differences can be of order unity.

Figure~\ref{diffdistrs} shows the difference $\Delta\chip$ between the averaged and heuristic estimates for the same populations of sources used in Fig.~\ref{manydistr}. The heuristic evaluation can either underestimate or overestimate the averaged result, with a distribution that is mildly skewed  toward  $\Delta\chip>0$. The largest disagreements depend on the injected value of $q$, $\chi_1$, and $\chi_2$ but can reach $\ssim 50\%$ for some of the cases shown in Fig.~\ref{diffdistrs}. As expected, differences are smaller for parameters that better satisfy the single-spin limit  $m_1^2\chi_1 \gg m_2^2\chi_2$.

Finally, let us briefly interpret our findings using the ``spin morphologies'' identified in Refs.~\cite{2015PhRvL.114h1103K,2015PhRvD..92f4016G}. These are mutually exclusive classes of sources where $|\Delta\Phi|$ can either circulate through full range $[0,\pi]$, librate about $|\Delta\Phi|=0$ and never reach $\pi$, or librate about $|\Delta\Phi|=\pi$ and never reach $0$. For $q=0.8$, $\chi_1=1$, {$\chi_2=1$}, $r\simeq 14 M$, and isotropic spins, we find that  $\Delta\chip= -0.06_{-0.27}^{+0.31}$ for binaries in the circulating morphology, $\Delta\chip= 0.46 _{-0.33}^{+0.27}$ for binaries that librate about $|\Delta\Phi|=0$, and  $\Delta\chip = -0.35_{-0.10}^{+0.30}$ for binaries that librate about $|\Delta\Phi|=\pi$ (where we indicated medians and 90\% interval). The heuristic estimate of $\chip$ assumes that contributions from $\Delta\Phi=0$ and $\Delta\Phi=\pi$ are equally important [cf. Eq.~(\ref{contributionaverage})]. This assumption is not  appropriate to describe binaries in the librating {morphologies} but provides a fairer description of the circulating sources. As shown in Ref.~\cite{2015PhRvD..92f4016G}, the number of sources in each of these three classes strongly depends on the binary parameters. %
The mismodeling  introduced by the heuristic definition of $\chip$ is therefore highly nonuniform,  affecting specific regions in the parameter space much {more} prominently than others.

\begin{table*}
\scalebox{0.97}{\centering
\begin{tabular}{l@{\hskip 0.2cm}|@{\hskip 0.2cm}cccc@{\hskip 0.3cm}||@{\hskip 0.3cm}l@{\hskip 0.2cm}|@{\hskip 0.2cm}cccc} 
{\bf Event} & $\chip^{\rm (Heu.)}$ & $\chip^{\rm (Asy.)}$ & $\chip^{\rm (Gen.)}$ & $\chip^{\rm (Av.)}$ &
{\bf Event} & $\chip^{\rm (Heu.)}$ & $\chip^{\rm (Asy.)}$ & $\chip^{\rm (Gen.)}$ & $\chip^{\rm (Av.)}$ \\
\hline \hline
GW150914 & 
$ 0.32^{+0.41}_{-0.26} $ &
$ 0.35^{+0.43}_{-0.28} $ &
$ 0.33^{+0.55}_{-0.27} $ &
$ 0.34^{+0.52}_{-0.27} $ &
GW190521\_074359 & 
$ 0.40^{+0.32}_{-0.29} $ &
$ 0.43^{+0.32}_{-0.31} $ &
$ 0.39^{+0.40}_{-0.30} $ &
$ 0.40^{+0.35}_{-0.28} $ \\
GW151012 & 
$ 0.31^{+0.40}_{-0.24} $ &
$ 0.33^{+0.42}_{-0.26} $ &
$ 0.31^{+0.51}_{-0.26} $ &
$ 0.33^{+0.45}_{-0.25} $ &
GW190527\_092055 & 
$ 0.46^{+0.42}_{-0.35} $ &
$ 0.49^{+0.43}_{-0.37} $ &
$ 0.47^{+0.56}_{-0.38} $ &
$ 0.48^{+0.48}_{-0.36} $ \\
GW151226 & 
$ 0.42^{+0.29}_{-0.27} $ &
$ 0.44^{+0.29}_{-0.28} $ &
$ 0.43^{+0.40}_{-0.31} $ &
$ 0.44^{+0.29}_{-0.27} $ &
GW190602\_175927 & 
$ 0.43^{+0.41}_{-0.32} $ &
$ 0.46^{+0.42}_{-0.34} $ &
$ 0.43^{+0.56}_{-0.34} $ &
$ 0.43^{+0.52}_{-0.32} $ \\
GW170104 & 
$ 0.37^{+0.36}_{-0.28} $ &
$ 0.40^{+0.36}_{-0.29} $ &
$ 0.38^{+0.48}_{-0.30} $ &
$ 0.39^{+0.41}_{-0.28} $ &
GW190620\_030421 & 
$ 0.43^{+0.37}_{-0.29} $ &
$ 0.46^{+0.37}_{-0.31} $ &
$ 0.44^{+0.48}_{-0.33} $ &
$ 0.45^{+0.43}_{-0.30} $ \\
GW170608 & 
$ 0.31^{+0.35}_{-0.24} $ &
$ 0.33^{+0.37}_{-0.25} $ &
$ 0.31^{+0.45}_{-0.25} $ &
$ 0.33^{+0.39}_{-0.25} $ &
GW190630\_185205 & 
$ 0.31^{+0.32}_{-0.23} $ &
$ 0.33^{+0.32}_{-0.24} $ &
$ 0.31^{+0.39}_{-0.24} $ &
$ 0.32^{+0.33}_{-0.23} $ \\
GW170729 & 
$ 0.42^{+0.36}_{-0.28} $ &
$ 0.45^{+0.37}_{-0.30} $ &
$ 0.43^{+0.50}_{-0.33} $ &
$ 0.44^{+0.44}_{-0.30} $ &
GW190701\_203306 & 
$ 0.42^{+0.42}_{-0.31} $ &
$ 0.45^{+0.42}_{-0.34} $ &
$ 0.43^{+0.53}_{-0.34} $ &
$ 0.43^{+0.51}_{-0.32} $ \\
GW170809 & 
$ 0.34^{+0.38}_{-0.26} $ &
$ 0.36^{+0.40}_{-0.28} $ &
$ 0.34^{+0.50}_{-0.28} $ &
$ 0.36^{+0.44}_{-0.27} $ &
GW190706\_222641 & 
$ 0.41^{+0.38}_{-0.28} $ &
$ 0.43^{+0.38}_{-0.30} $ &
$ 0.42^{+0.49}_{-0.32} $ &
$ 0.43^{+0.44}_{-0.30} $ \\
GW170814 & 
$ 0.49^{+0.32}_{-0.37} $ &
$ 0.52^{+0.33}_{-0.40} $ &
$ 0.49^{+0.52}_{-0.39} $ &
$ 0.49^{+0.48}_{-0.38} $ &
GW190707\_093326 & 
$ 0.29^{+0.39}_{-0.23} $ &
$ 0.31^{+0.40}_{-0.24} $ &
$ 0.28^{+0.47}_{-0.23} $ &
$ 0.30^{+0.40}_{-0.24} $ \\
GW170818 & 
$ 0.51^{+0.29}_{-0.35} $ &
$ 0.54^{+0.30}_{-0.36} $ &
$ 0.51^{+0.47}_{-0.39} $ &
$ 0.51^{+0.43}_{-0.35} $ &
GW190708\_232457 & 
$ 0.29^{+0.43}_{-0.24} $ &
$ 0.31^{+0.44}_{-0.25} $ &
$ 0.28^{+0.45}_{-0.23} $ &
$ 0.31^{+0.39}_{-0.24} $ \\
GW170823 & 
$ 0.45^{+0.41}_{-0.35} $ &
$ 0.48^{+0.42}_{-0.37} $ &
$ 0.46^{+0.57}_{-0.37} $ &
$ 0.47^{+0.52}_{-0.35} $ &
GW190720\_000836 & 
$ 0.33^{+0.43}_{-0.22} $ &
$ 0.35^{+0.45}_{-0.24} $ &
$ 0.34^{+0.50}_{-0.25} $ &
$ 0.35^{+0.44}_{-0.23} $ \\
GW190408\_181802 & 
$ 0.39^{+0.37}_{-0.31} $ &
$ 0.42^{+0.39}_{-0.33} $ &
$ 0.38^{+0.46}_{-0.31} $ &
$ 0.39^{+0.39}_{-0.30} $ &
GW190727\_060333 & 
$ 0.48^{+0.39}_{-0.36} $ &
$ 0.51^{+0.40}_{-0.38} $ &
$ 0.49^{+0.57}_{-0.40} $ &
$ 0.50^{+0.51}_{-0.37} $ \\
GW190412 & 
$ 0.31^{+0.19}_{-0.16} $ &
$ 0.32^{+0.19}_{-0.16} $ &
$ 0.31^{+0.23}_{-0.20} $ &
$ 0.32^{+0.19}_{-0.16} $ &
GW190728\_064510 & 
$ 0.29^{+0.37}_{-0.20} $ &
$ 0.31^{+0.39}_{-0.21} $ &
$ 0.29^{+0.44}_{-0.22} $ &
$ 0.31^{+0.38}_{-0.21} $ \\
GW190413\_052954 & 
$ 0.42^{+0.42}_{-0.32} $ &
$ 0.45^{+0.43}_{-0.33} $ &
$ 0.42^{+0.55}_{-0.33} $ &
$ 0.43^{+0.49}_{-0.32} $ &
GW190731\_140936 & 
$ 0.42^{+0.43}_{-0.32} $ &
$ 0.45^{+0.45}_{-0.34} $ &
$ 0.42^{+0.58}_{-0.34} $ &
$ 0.43^{+0.51}_{-0.32} $ \\
GW190413\_134308 & 
$ 0.56^{+0.36}_{-0.42} $ &
$ 0.60^{+0.36}_{-0.45} $ &
$ 0.59^{+0.52}_{-0.47} $ &
$ 0.58^{+0.48}_{-0.43} $ &
GW190803\_022701 & 
$ 0.45^{+0.42}_{-0.34} $ &
$ 0.47^{+0.43}_{-0.36} $ &
$ 0.45^{+0.58}_{-0.36} $ &
$ 0.46^{+0.54}_{-0.34} $ \\
GW190421\_213856 & 
$ 0.49^{+0.40}_{-0.37} $ &
$ 0.52^{+0.40}_{-0.39} $ &
$ 0.49^{+0.56}_{-0.39} $ &
$ 0.50^{+0.52}_{-0.37} $ &
GW190814 & 
$ 0.04^{+0.04}_{-0.03} $ &
$ 0.05^{+0.04}_{-0.03} $ &
$ 0.04^{+0.05}_{-0.03} $ &
$ 0.05^{+0.04}_{-0.03} $ \\
GW190424\_180648 & 
$ 0.52^{+0.37}_{-0.37} $ &
$ 0.55^{+0.38}_{-0.40} $ &
$ 0.54^{+0.57}_{-0.42} $ &
$ 0.54^{+0.52}_{-0.39} $ &
GW190828\_063405 & 
$ 0.43^{+0.36}_{-0.30} $ &
$ 0.46^{+0.36}_{-0.32} $ &
$ 0.44^{+0.48}_{-0.34} $ &
$ 0.45^{+0.44}_{-0.32} $ \\
GW190503\_185404 & 
$ 0.39^{+0.41}_{-0.29} $ &
$ 0.41^{+0.42}_{-0.31} $ &
$ 0.39^{+0.52}_{-0.31} $ &
$ 0.40^{+0.47}_{-0.30} $ &
GW190828\_065509 & 
$ 0.29^{+0.40}_{-0.22} $ &
$ 0.30^{+0.40}_{-0.23} $ &
$ 0.29^{+0.43}_{-0.24} $ &
$ 0.30^{+0.39}_{-0.23} $ \\
GW190512\_180714 & 
$ 0.23^{+0.37}_{-0.18} $ &
$ 0.24^{+0.38}_{-0.19} $ &
$ 0.23^{+0.39}_{-0.19} $ &
$ 0.24^{+0.34}_{-0.18} $ &
GW190910\_112807 & 
$ 0.41^{+0.39}_{-0.32} $ &
$ 0.44^{+0.41}_{-0.34} $ &
$ 0.40^{+0.49}_{-0.32} $ &
$ 0.40^{+0.45}_{-0.31} $ \\
GW190513\_205428 & 
$ 0.30^{+0.40}_{-0.22} $ &
$ 0.32^{+0.41}_{-0.24} $ &
$ 0.31^{+0.48}_{-0.25} $ &
$ 0.32^{+0.40}_{-0.24} $ &
GW190915\_235702 & 
$ 0.56^{+0.36}_{-0.39} $ &
$ 0.59^{+0.35}_{-0.41} $ &
$ 0.57^{+0.50}_{-0.44} $ &
$ 0.57^{+0.43}_{-0.40} $ \\
GW190514\_065416 & 
$ 0.47^{+0.39}_{-0.34} $ &
$ 0.50^{+0.40}_{-0.37} $ &
$ 0.47^{+0.59}_{-0.38} $ &
$ 0.47^{+0.55}_{-0.34} $ &
GW190924\_021846 & 
$ 0.24^{+0.40}_{-0.18} $ &
$ 0.26^{+0.42}_{-0.20} $ &
$ 0.24^{+0.45}_{-0.19} $ &
$ 0.26^{+0.40}_{-0.19} $ \\
GW190517\_055101 & 
$ 0.48^{+0.31}_{-0.28} $ &
$ 0.52^{+0.31}_{-0.30} $ &
$ 0.50^{+0.46}_{-0.36} $ &
$ 0.50^{+0.40}_{-0.31} $ &
GW190929\_012149 & 
$ 0.39^{+0.43}_{-0.30} $ &
$ 0.41^{+0.42}_{-0.31} $ &
$ 0.41^{+0.46}_{-0.33} $ &
$ 0.41^{+0.42}_{-0.31} $ \\
GW190519\_153544 & 
$ 0.45^{+0.34}_{-0.29} $ &
$ 0.48^{+0.33}_{-0.31} $ &
$ 0.46^{+0.46}_{-0.34} $ &
$ 0.46^{+0.39}_{-0.30} $ &
GW190930\_133541 & 
$ 0.34^{+0.40}_{-0.24} $ &
$ 0.36^{+0.41}_{-0.25} $ &
$ 0.34^{+0.48}_{-0.26} $ &
$ 0.35^{+0.40}_{-0.25} $ \\
GW190521 & 
$ 0.67^{+0.26}_{-0.44} $ &
$ 0.72^{+0.26}_{-0.46} $ &
$ 0.70^{+0.58}_{-0.52} $ &
$ 0.70^{+0.56}_{-0.46} $ &

\end{tabular}
}
\caption{Medians and 90\% confidence intervals of $\chip$ for 45 GW events from the first three observing runs of the LIGO/Virgo interferometers. For each event, we consider all four definitions of $\chip$:  heuristic [Eq.~(\ref{usualchip})],   asymptotic [Eq.~(\ref{ellipE})], generalized [Eq.~(\ref{fullchip})], and averaged [Eq.~(\ref{fullaverage})].}
\label{bigtable}
\end{table*}

\section{Impact on current LIGO events}
\label{currentevents}

The examples presented so far indicate that a consistent generalization of $\chip$ can in principle be an important player in the interpretation of BH-binary  observations. We now turn our attention to current GW events from the first three observing runs of the LIGO/Virgo detectors. 

For O1 and O2, we make use of publicly released posterior samples from Ref.~\cite{2020MNRAS.499.3295R}, which include all 10 BH binary events reported in the GWTC-1 catalog\footnote{We could not use the posterior samples publicly  released with Ref.~\cite{2019PhRvX...9c1040A} because they do not include the variable $\Delta\Phi$. The analyses of Refs.~\cite{2020MNRAS.499.3295R} and \cite{2019PhRvX...9c1040A} were found to be compatible.} \cite{2019PhRvX...9c1040A}. For O3a, we use data products released together with the GWTC-2 catalog \cite{2020arXiv201014527A} and consider all BH binary events with false-alarm rate $<$1/yr. The resulting sample of 45  detections is reported in Table~\ref{bigtable}.

The O1-O2 analysis of Ref.~\cite{2020MNRAS.499.3295R} employs the IMRPhenomPv2 \cite{2014PhRvL.113o1101H} waveform model, where two-spin effects are not fully included. For the case of GW151226, we cross-checked our results using posterior samples obtained with the more accurate IMRPhenomPv3 model from Ref.~\cite{2019PhRvD.100b4059K} and did not detect significant differences. Unless specified otherwise, events from O3a are analyzed using combined samples obtained with different waveform families as described in Appendix A.1 of Ref.~\cite{2020arXiv201014527A}.
LIGO/Virgo parameter estimation samples report the spin directions at a fixed GW frequency, which was set to $f_{\rm ref} = 20$ Hz for all the events but GW190521 where the high mass imposed a lower value $f_{\rm ref}=11$ Hz. We convert GW frequency to PN separation $r$  as described in Appendix~\ref{averimp}.%

\begin{figure*}[p]
\includegraphics[width=\textwidth]{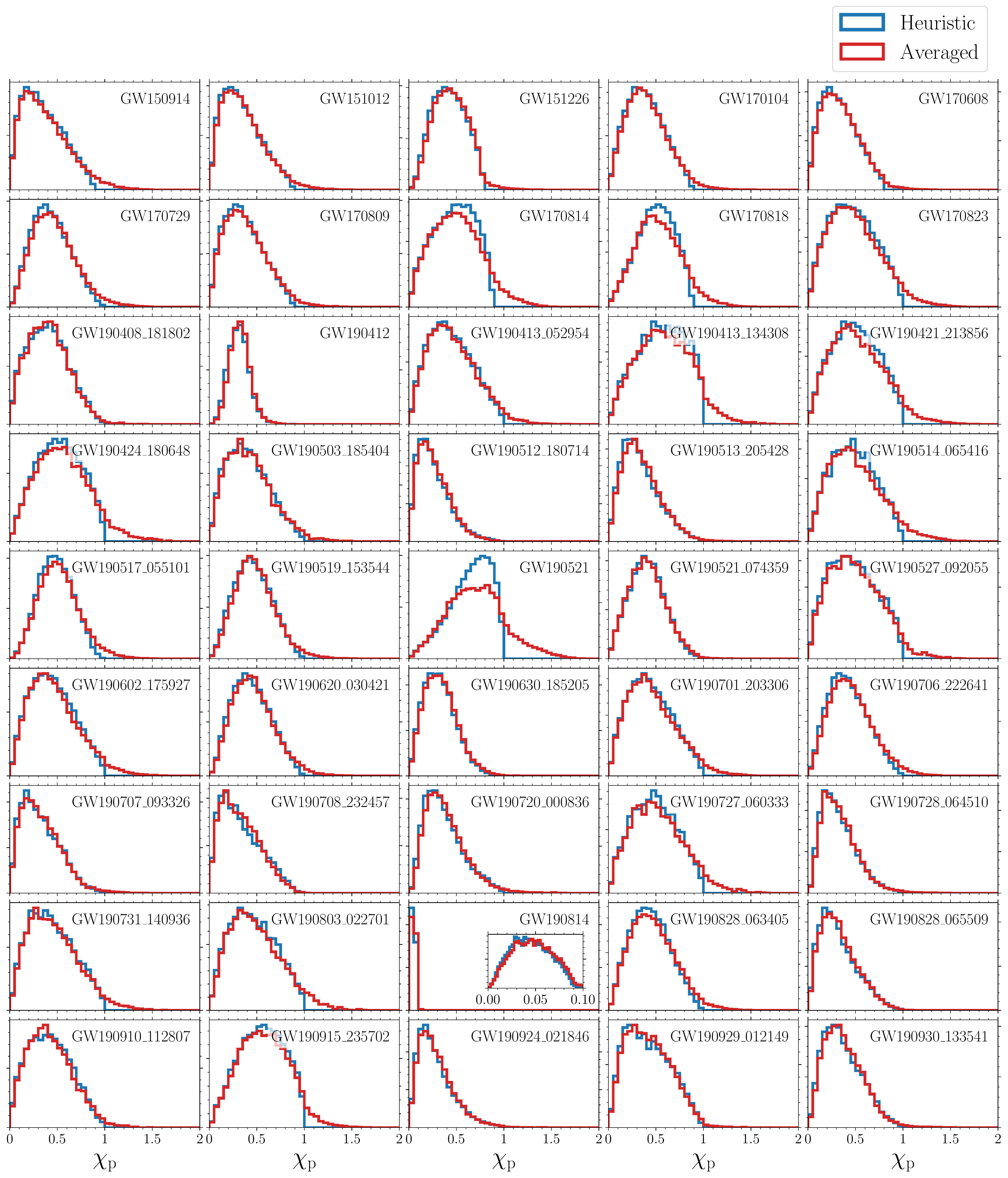}
\caption{Posterior distributions for the heuristic  [Eq.~(\ref{usualchip}), blue] and averaged  [Eq.~(\ref{fullaverage}), red] estimate of $\chip$ for 45 GW events from the first, second, and third LIGO/Virgo observing runs. Posteriors for the asymptotic (generalized) $\chip$ are qualitatively similar to the heuristic (averaged) results and are omitted for clarity. We report the generic tendency of tails extending toward the region where $\chip\gtrsim 1$.}
\label{allevents}
\end{figure*}

Figure \ref{allevents} shows posterior distributions for the heuristic (blue) and averaged (red) estimates of $\chip$ for all 45 events. Medians and symmetric  90\% confidence intervals are reported in Table~\ref{bigtable}. For most events, the posterior distributions obtained with different definitions of $\chip$ are qualitatively similar, indicating that the generalization proposed in this paper does not alter the physical interpretation of these systems, at least at the present {SNR}. However, our $\chip$ distributions present an overall tendency toward larger values. This appears to be a rather generic feature: the posteriors of the averaged and generalized $\chip$'s have longer tails compared to both the heuristic and asymptotic ones. Some cases that are worth singling out from Fig.~\ref{allevents} are %
GW170814,
GW170823,
GW190413\_134308,
GW190421\_213856,
GW190424\_180648,
GW190514\_065416,
GW190521 (see below),
GW190727\_060333,
GW190803\_022701, and
GW190915\_235702.

In these cases, the posterior of the heuristic $\chip$ is somewhat steep near its $\chip=1$ boundary. On the other hand, the averaged and generalized distributions extend smoothly into the $\chip\gtrsim 1$ region. The heuristic definition of $\chip$ from Eq.~(\ref{usualchip}) causes an artificial pileup of posterior samples at the boundary $\chip\lesssim 1$; the samples affected are those with in-plane spin components which are moderately large and coaligned.

\begin{figure}
\includegraphics[width=\columnwidth]{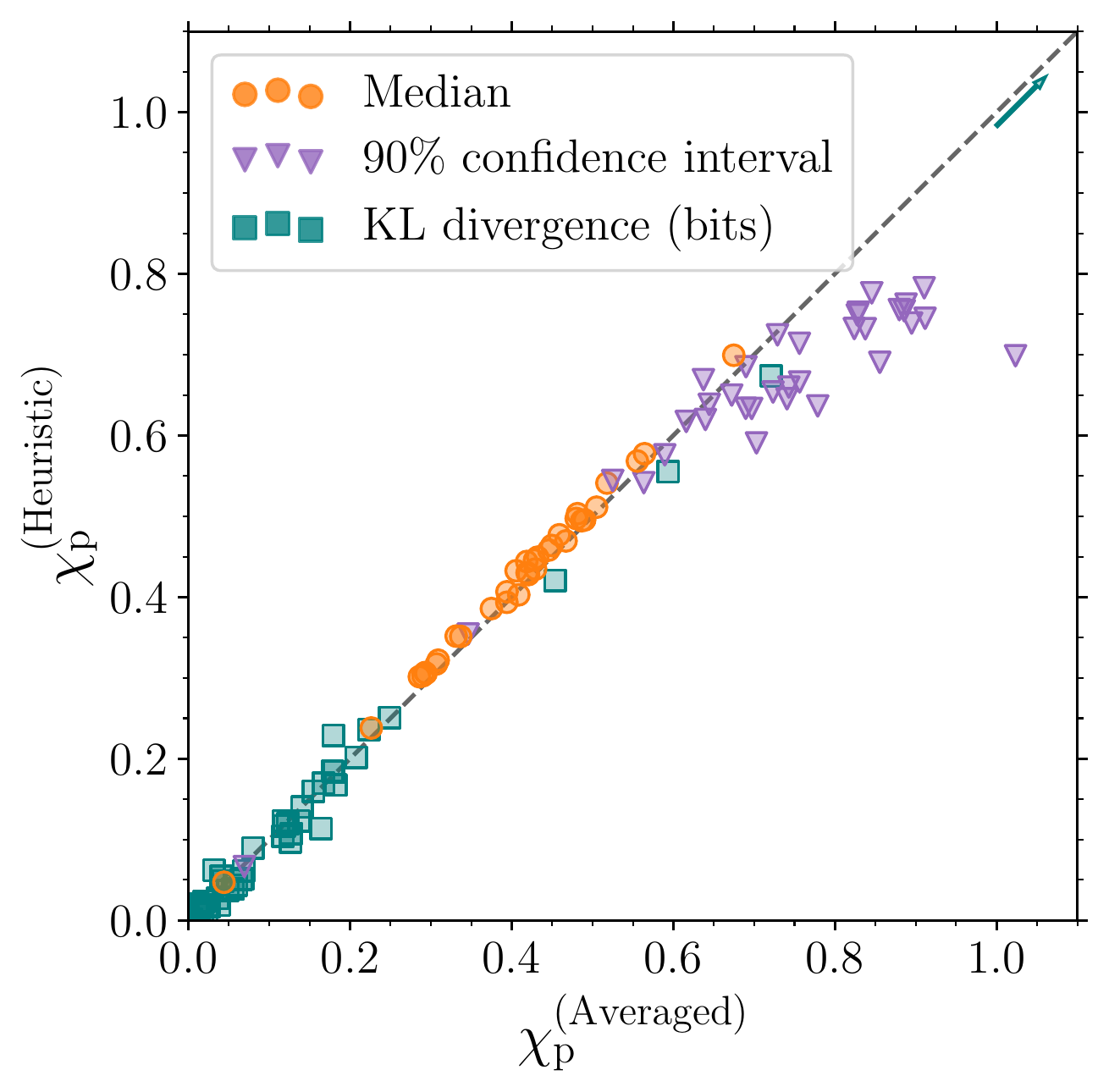}
\caption{Statistical properties of the precession parameter $\chip$ for current GW events. We contrast the averaged [Eq.~(\ref{fullaverage}), $x$-axis] and heuristic [Eq.~(\ref{usualchip}), $y$-axis]  definitions of $\chip$. Scatter points show the medians of the posterior distributions (orange circles), the width of their 90\% confidence interval (purple triangles), and the KL divergence between prior and posterior measured in bits (teal squares).
The KL divergence of GW190814 is $\ssim 4.3$ bits, which is off the scale of this figure in the direction of the arrow.}
\label{scatterLIGO}
\end{figure}

\begin{figure*}[t]
\centering
\includegraphics[width=\textwidth]{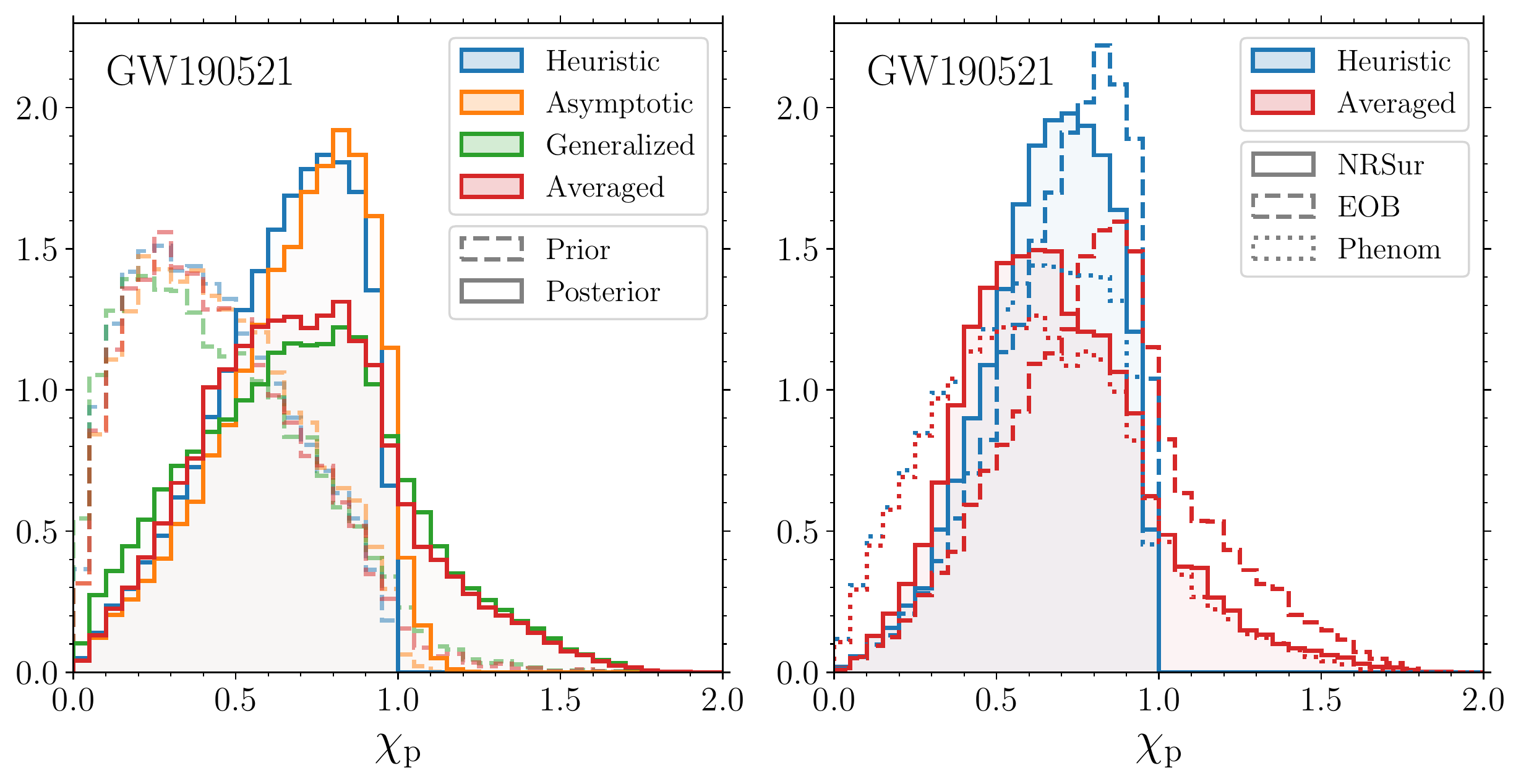}
\caption{Precession parameter $\chip$ for GW190521. Colors indicate the four definitions of $\chip$ described in this paper: heuristic [Eq.~(\ref{usualchip}), blue],   asymptotic [Eq.~(\ref{ellipE}), orange], generalized [Eq.~(\ref{fullchip}), green], and averaged [Eq.~(\ref{fullaverage}), red]. The left panel shows prior (dashed) and posterior (solid) distributions obtained by combining samples from different waveform approximants. The right panel shows posterior distributions obtained with three different waveform models: NRSur7dq4 (``NRsur'', solid), SEOBNRv4PHM  (``EOB'', dashed), and IMRPhenomPv3HM (``Phenom'', dotted).}
\label{only190521}
\end{figure*}

Some statistical properties of these $\chip$ measurements are summarized in Fig.~\ref{scatterLIGO} where we contrast the heuristic and the averaged definitions. At least at the current detector sensitivity, medians of the $\chip$ posteriors (orange circles) are not sensitive to the generalizations put forward in this paper. On the other hand, the width of their 90\% confidence interval %
depends  on the $\chip$ definition. In particular, we find that the common heuristic approach systematically underestimates the $\chip$ measurement errors. This is once more  due to the  $\chip\gtrsim 1$ tails of the posteriors shown in Fig.~\ref{allevents}, which push the upper edge of the 90\% confidence interval toward larger values. As shown in Fig.~\ref{scatterLIGO}, the majority of the current GW events are affected, with potential consequences for current and future population studies. 

Figure~\ref{scatterLIGO} also shows the Kullback-Leibler (KL) divergence $D_{\rm KL}$ between prior and posterior samples evaluated using the averaged and heuristic estimates of $\chip$. The $\chip$ KL divergence, also known as relative entropy, is often used as a metric to discriminate whether the data contain enough evidence for spin precession. For the few systems with $D_{
\rm KL}\gtrsim 0.4$ bits (GW190412, GW190512\_180714, GW190521, and GW190814), the relative entropy of the averaged $\chip$ is up to $7\%$ higher, indicating that current estimates tend to mildly underestimate the information gain. %

In terms of the alternative metric $\rho_{\rm p}$ \cite{2020PhRvD.102b4055F,2020PhRvD.102d1302F,2020arXiv201004131G}, Ref.~\cite{2020arXiv201014527A} reports that the events with the largest excess SNR are GW190412 ($\rho_{\rm p}=3.0$) and GW190521 ($\rho_{\rm p}=1.6$), which are also among those we highlight. Posterior distributions of $\rho_{\rm p}$ for some of the current events are reported in  Fig.~1 of Ref.~\cite{2020PhRvD.102d1302F} and Fig.~1 of Ref.~\cite{2020arXiv201004131G}, and can be compared against our Fig.~\ref{allevents}.

The most striking case from Fig.~\ref{allevents} is undoubtedly that of GW190521. Classified as the most massive event to date, GW190521 is a BH binary with total mass $\ssim 150 M_\odot$ which shows some preference for spinning, precessing  BHs (odds ratio of $\ssim 10:1$) \cite{2020PhRvL.125j1102A}. The $\chip$ properties of this event are singled out in Fig.~\ref{only190521}. The left panel shows prior and posterior distributions for all four definitions of $\chip$. While the priors are all qualitatively similar, the posteriors (here computed averaging over different waveform families \cite{2020arXiv201014527A}) show more pronounced differences between the heuristic/asymptotic and the generalized/averaged $\chip$'s. The $\chip$ generalization put forward in this paper affects the posterior much more prominently than the prior, and thus highlights features which are present in the data. The right panel shows posterior distributions obtained using three different waveform models ---NRSur7dq4 \cite{2019PhRvR...1c3015V}, SEOBNRv4PHM \cite{2020PhRvD.102d4055O}, and IMRPhenomPv3HM \cite{2020PhRvD.101b4056K}--- which all include two-spin effects  and higher-order modes. While some dependence on the waveform  is present (cf.~\cite{2020arXiv201014527A}), we find that the differences between the definitions of $\chip$ cannot be absorbed within these uncertainties. At least for GW190521, the generalization of $\chip$  presented here dominates over waveform systematics.

\section{Conclusions}
\label{concl}

The parameter $\chip$ is commonly used to characterize relativistic precession in GW observations of BH binaries. The reduction of the spin-precession problem to a single parameter was initially motivated by efficient waveform construction and template placing \cite{2015PhRvD..91b4043S}. The popular waveform model of Ref.~\cite{2014PhRvL.113o1101H} indeed makes direct use of a single effective precessing spin. 
Waveform models have evolved since then and now fully include two-spin effects~\cite{2019PhRvR...1c3015V,2020PhRvD.102d4055O,2020PhRvD.101b4056K}. Although GW parameter-estimation algorithms sample all six Cartesian components of the BH spins, information on the spin components perpendicular to the orbital angular momentum is often condensed into $\chip$ for interpretation purposes (for measurement accuracies on the individual spins see e.g.~\cite{2014PhRvL.112y1101V,2016PhRvD..93h4042P}).
 It is indeed very desirable to have a single parameter that, if measured confidently, can be interpreted as ``the amount of precession'' in a given GW observation.

The parameter $\chip$ was first defined in Ref.~\cite{2015PhRvD..91b4043S} with specific assumptions that are here relaxed for the first time. In particular, we propose that the %
common  definition
\begin{align}
\chip =  \max\left(\chi_1 \sin\theta_1, q\frac{4q + 3}{4 + 3q} \chi_2 \sin\theta_2\right)
\end{align}
should be generalized to
\begin{align}
\chip &= \bigg[\left(\chi_1 \sin\theta_1 \right)^2 +   \left(q\frac{4q + 3}{4 + 3q} \chi_2 \sin\theta_2 \right)^2
\notag \\
&\quad+ 2  q\frac{4q + 3}{4 + 3q} \chi_1\chi_2 \sin\theta_1 \sin\theta_2 \cos\Delta\Phi \bigg]^{1/2}\,.
\label{chipagain}
\end{align}
The latter can then be precession averaged as in Eq.~(\ref{fullaverage}) and Appendix~\ref{averimp}. For a public implementation using the {\sc Python} programming language see  \href{https://github.com/dgerosa/generalizedchip}{github.com/dgerosa/generalizedchip} \cite{generalizedchip}.

The crucial difference between the two definitions above is that $\chip$ depends not only on the magnitudes of the in-plane spin components $\chi_1\sin\theta_1$ and $\chi_2\sin\theta_2$ but also on the angle $\Delta\Phi$ between them. 

It is worth noting that the generalization  we propose is bound by $\chip
\leq 2$, compared to $\chip\leq 1$ for the heuristic definition. This reflects one's intuition that binaries where both BHs contribute significantly to the precession dynamics cannot be reduced to an effective system with a single spin.
From the definition of  Eq.~(\ref{chipagain}) one can immediately prove that $\chip<1$ if either spin is parallel to the orbital angular momentum ($\chi_i\sin\theta_i=0$). It follows that the additional region $\chip>1$ is 
\emph{exclusive} to binaries with two precessing spins. Much like an observation where $\chip$ is confidently $>0$ would indicate the presence of at least one precessing spin, a GW event in the $\chip>1$ region can be interpreted as a detection of two-spin effects.

It is important to note that there is some arbitrariness in the precise definition of $\chip$.  For instance, instead of the magnitude $|{d\hat{\vec{L}}}/{dt}|$ adopted in Sec.~\ref{maths}, one could use the projection of the total spin onto the orbital plane
\begin{align}
\chi_\perp &\equiv \frac{|(\vec{S}_1 + \vec{S_2}) \cross \hat{\vec{L}}|}{M^2} =  \frac{1}{(1+q)^2}  \big[\left(\chi_1 \sin\theta_1 \right)^2 
\notag \\
& +\left(q \chi_2 \sin\theta_2 \right)^2
+ 2  q \chi_1\chi_2 \sin\theta_1 \sin\theta_2 \cos\Delta\Phi \big]^{1/2}\,,
\end{align}
which differs from Eq.~(\ref{chipagain}) only by some factors of $q$ (see also~\cite{2020arXiv201202209T}).  Similarly, in Eq.~(\ref{fullchip}) we, somehow arbitrarily, opted for normalizing the magnitude of ${d\hat{\vec{L}}}/{dt}$ by the precession frequency of the primary BH $\Omega_1$. This is the same choice made in Ref.~\cite{2015PhRvD..91b4043S} and was here retained to allow for a meaningful comparison between our results and theirs. 
This ensures that our $\chip$ re-definition agrees with the heuristic one in the $\chip\to 0$ limit,  as evidenced by the small-$\chip$ regions in Fig.~\ref{allevents}. A reflection of this feature is that the single-spin limit is preserved  [cf. Eq.~(\ref{tayloraverage})].
An alternative normalization, which goes further in the direction of putting the two BHs on equal footing, would be to divide $|{d\hat{\vec{L}}}/{dt}|$ by $\Omega_1+\Omega_2$. Our results can trivially be rescaled to that choice by the transformation
\begin{equation}
\chip\; \longrightarrow \; \frac{\chip}{1+\tilde \Omega}\;.
\end{equation}
In this case, one would obtain a precession parameter that is $\leq 1$ but  it would present  a different small-spin behavior, resulting in an estimator that cannot be easily compared with the heuristic definition. %

We stress that our recipe for evaluating $\chip$ does not require new or different parameter-estimation runs, which are computationally expensive, but can be carried out entirely in postprocessing. In this paper, we pursued this strategy using public posterior samples from the LIGO/Virgo events reported to date. We report the generic occurrence  of long tails in the posterior distributions of $\chip$ that extend smoothly into the previously forbidden region where $\chip \gtrsim 1$.  The most relevant case to date which shows the importance of defining a consistent precession parameter is GW190521, where  $p(\chip>1)\simeq  16\%$. This number can be interpreted as a lower limit to the probability that GW190521 contained two precessing spins.  In total, there are 6 (29) events for which this ``two-spin probability'' is greater than $5\%$ ($1\%$). %

Such tails in the $\chip$ posteriors have two main consequences:
\begin{itemize}
\item First, they enlarge the 90\% confidence interval of $\chip$, indicating that current estimates of the measurement errors with which precession is measured might be underestimated.
\item Second, more posterior weight is placed at higher values of $\chip$, which can  arguably be interpreted as an indication that the data show more evidence for spin precession than previously reported.
\end{itemize}

Both these points might have important consequences for the astrophysical interpretation of GW events, because spin precession is thought to be a key tracer of BH binary formation pathways. 

The issues of constructing waveform models and that of interpreting observations are not decoupled. The relative success of approximants which rely on effective-spin  parameters indicates that two-spin effects are subdominant and intrinsically difficult to measure~\cite{2014PhRvL.112y1101V,2016PhRvD..93h4042P}. One should also keep in mind that the event posteriors are dependent on the waveform models used in the analysis and are therefore subject to their systematics \cite{2020arXiv201014527A}. With these caveats in mind, our results indicate that some of the current events present some hints of two-spin physics. Data are bound to become more informative as detectors improve in sensitivity.

Current GW population studies (e.g. Ref.~\cite{2020arXiv201014533T}) all make use of $\chip$ at a fixed reference frequency $f_{\rm ref}$ and neglect its evolution along the inspiral. This issue is exacerbated by the common adoption of the heuristic expression of $\chip$, which varies on the short timescale of the problem. The averaging procedure proposed here is the most natural mitigation strategy.

Large sets of software injections are necessary to better understand how these augmented $\chip$ estimators respond to current LIGO/Virgo parameter-estimation pipelines and compare with other metrics such as excess SNR $\rho_{\rm p}$ and Bayes' factor. In this paper, 
we have only investigated the relevance of our redefinition of $\chip$ on individual events, not its collective effect on the    detected population. The impact of our findings on population studies might be significant because the generalization of $\chip$ here proposed affects current events in a weak but systematic manner.

\acknowledgements

We thank  
Vishal Baibhav,
Emanuele Berti,
Riccardo Buscicchio, 
Michael Kesden, 
Christopher Moore, 
Richard O'Shaughnessy, 
Isobel Romero-Shaw,
Nathan Steinle, 
Evangelos Stoikos, 
and
Daniel Wysocki 
for discussions, constructive comments, and technical help. We thank the LIGO/Virgo Collaboration and the authors of Ref.~\cite{2020MNRAS.499.3295R} for their public data releases. 
 We thank Katerina Chatziioannou and the authors of Ref.~\cite{2019PhRvD.100b4059K} for sharing their posterior samples with us.
D.Ger., M.M. and D.Gan. are supported by European Union H2020 ERC Starting Grant No. 945155--GWmining and
Royal Society Grant No. RGS-R2-202004. D.Ger. is supported by Leverhulme Trust Grant No. RPG-2019-350. 
M.M., D.Gan., and L.M.T. are supported by the STFC and the School of Physics and Astronomy at the University of Birmingham. P.S. is supported by Dutch Research Council (NWO) Veni Grant No. 680-47-460.
Computational work was performed on the University of Birmingham BlueBEAR cluster, the Athena cluster at HPC Midlands+ funded by EPSRC Grant No. EP/P020232/1, and the Maryland Advanced Research Computing Center (MARCC).

\appendix
\section{Precession-average implementation}
\label{averimp}

In this Appendix we provide a practical implementation of  Eq.~(\ref{fullaverage}). We rely on the multi-timescale framework presented in Refs.~\cite{2015PhRvL.114h1103K,2015PhRvD..92f4016G, 2017PhRvD..95j4004C}, where precession cycles are parametrized using the magnitude of the total spin  $S(t)=|\vec{S}_1(t)+\vec{S}_2(t)|$.

Current LIGO/Virgo pipelines provide the binary configurations in terms of detector-frame total mass $M$, mass ratio $q$, spin magnitude $
\chi_{1}$, $\chi_2$, and the orientations $\theta_1$, $\theta_2$, $\Delta\Phi$ at a given GW frequency $f_{\rm ref}$.  
First, we obtain the corresponding separation $r$ using the PN expression reported in  Eq. (4.13) of Ref.~\cite{1995PhRvD..52..821K},
\begin{align}
\frac{r}{M} &=  (M \omega)^{-2/3} - \left[ 1- \frac{q}{3 (1+q)^2} \right]
\notag\\
&- \frac{(M \omega)^{1/3}}{3 (1+q)^2} \left[(3q+2)\chi_1\cos\theta_1 + q(3+2q)\chi_2 \cos\theta_2 \right]
\notag\\
&+
(M \omega)^{2/3}
\bigg\{\frac{q}{(1+q)^2} \left[ \frac{19}{4} + \frac{q}{9(1+q)^2}\right]
\notag\\
&-\frac{\chi_1\chi_2}{2} (\sin\theta_1\sin\theta_2\cos\Delta\Phi - 2 \cos\theta_1\cos\theta_2)
\bigg\} \, ,
\end{align}
where $\omega=\pi f_{\rm ref}$ is the orbital angular velocity.

One can then evaluate quantities that are constant on the precession timescale. These include the magnitudes of the spins $S_{i} = \chi_i m_i^2$ ,  the magnitude of the Newtonian angular momentum $L = m_1 m_2 \sqrt{r/M}$, the effective spin $\chi_{\rm eff}$ from  Eq.~(\ref{chieff}), and the magnitude of the total angular momentum 
\begin{align}
J&= [ L^2 + S_1^2+S_2^2 +2L (S_1 \cos\theta_1 + S_2\cos\theta_2) 
\notag \\
&+ 2 S_1 S_2 (\sin\theta_1\sin\theta_2\cos\Delta\Phi + \cos\theta_1\cos\theta_2)]^{1/2}\,.
\label{Jeq}
\end{align}

These quantities can be {used} to derive the following parametric expressions of spin angles as a function of $S${,}
\begin{align}
\cos\theta_1(S) &= \frac{1}{2 (1-q) S_1} \left[ \frac{J^2-L^2-S^2}{L} - \frac{2 q M^2 \chieff}{1+q}\right]\,,
\\
\cos\theta_2(S) &=\frac{q}{2 (1-q) S_2} \left[ -\frac{J^2-L^2-S^2}{L} + \frac{2  M^2 \chieff}{1+q}\right]\,,
\\
\cos\Delta\Phi(S) & =\frac{S^2 - S_1^2 - S_2^2 -2 S_1 S_2 \cos\theta_1(S) \cos\theta_2(S)}{ 2 S_1 S_2 \sin\theta_1(S) \sin\theta_2(S)} \,,
\end{align}
which can be substituted into Eq.~(\ref{fullchip}) to obtain $\chip(S)$.

The time derivative of $S$ follows directly from the spin-precession equations (e.g. \cite{2008PhRvD..78d4021R}):
\begin{align}
\left| \frac{d S}{dt}\right |  &= \frac{3}{2}  \frac{S_1 S_2 M^9}{L^5} \frac{q^5(1-q)}{(1+q)^{11}} \left[ 1 - \frac{qM^2 \chieff}{(1+q)^2 L } \right]
\notag \\
&\times \frac{\sin\theta_1(S)\sin\theta_2(S) |\sin\Delta\Phi(S)|}{S}\,.
\label{dSdt}
\end{align}
One can prove \cite{2015PhRvD..92f4016G} that there are two stationary points where ${d S}/{dt} = 0$, which we refer to as $S_\pm$ with $S_-\leq S_+$.  The absolute values in Eqs.~(\ref{dSdt}) are related to the fact that the two halves of the precession cycle $S_-\to S_+$ and $S_+\to S_-$ are symmetric, so we can safely integrate only over the first half.  

Putting  all these ingredients together yields
 \begin{align}
\langle\chip \rangle = \frac{\displaystyle \bigintssss_{S_-}^{S_+} \chip(S)\left| \frac{d S}{dt}\right |^{-1} dS}{  \displaystyle\bigintssss_{S_-}^{S_+} \left| \frac{d S}{dt}\right |^{-1} \!\!\!\! dS}\,.
\label{avchipS}
\end{align}

\noindent
Parametrizing the precession cycle in terms of $S(t)$ holds as long as $q<1$. The angle $\varphi'$ of Eq.~(\ref{varphiprime}) should instead be used if $q=1$ (which, strictly speaking, never happens)~\cite{2017CQGra..34f4004G}.

\bibliography{generalizedchip}

\end{document}